\documentclass[aps,prd,twocolumn,epsf,superscriptaddress]{revtex4}
\newcommand{\postscript}[2]{\setlength{\epsfxsize}{#2\hsize}
   \centerline{\epsfbox{#1}}}

\usepackage{amsfonts}
\usepackage{amsmath}
\usepackage{amssymb,amsbsy}
\usepackage{bm}
\usepackage{makeidx}
\usepackage{latexsym}
\usepackage{graphicx}
\usepackage{epsfig}
\usepackage{subfigure}
\usepackage{dsfont}
\usepackage{mathcomp}
\usepackage{color}
\usepackage{mathrsfs}

\newcommand{\be}{\begin{equation}}
\newcommand{\ee}{\end{equation}}
\newcommand{\bea}{\begin{eqnarray}}
\newcommand{\eea}{\end{eqnarray}}

\newcommand{\gapp}{\mathrel{\raise.3ex\hbox{$>$}\mkern-14mu
              \lower0.6ex\hbox{$\sim$}}}
\newcommand{\lapp}{\mathrel{\raise.3ex\hbox{$<$}\mkern-14mu
              \lower0.6ex\hbox{$\sim$}}}

\begin{document}
\title{Vanishing Dimensions and Planar Events at the LHC}
\author{Luis Anchordoqui}
\affiliation{Department of Physics, University of Wisconsin-Milwaukee, Milwaukee, WI 53201, USA}

\author{De Chang Dai}

\affiliation{HEPCOS, Department of Physics, SUNY at Buffalo, Buffalo, NY 14260-1500, USA}
\affiliation{Astrophysics, Cosmology and Gravity Centre, University of Cape Town, Rondebosch, Private Bag, 7700, South Africa}

\author{Malcolm Fairbairn}

\affiliation{Department of Physics, Kings College London, Strand, London WC2R 2LS, UK}

\author{Greg Landsberg}
\affiliation{Department of Physics, Brown University, Providence,
RI 02912, USA}

\author{Dejan Stojkovic}
\affiliation{HEPCOS, Department of Physics,
SUNY at Buffalo, Buffalo, NY 14260-1500, USA}

\begin{abstract}
\noindent
We propose that the effective dimensionality of the space we live in depends on the length scale we are probing. As the length scale increases, new dimensions open up. At short scales the space is lower dimensional; at the intermediate scales the space is three-dimensional; and at large scales, the space is effectively higher dimensional. This setup allows for some fundamental problems in cosmology, gravity, and particle physics to be attacked from a new perspective. The proposed framework, among the other things, offers a new approach to the cosmological constant problem and results in striking collider phenomenology and may explain elongated jets observed in cosmic-ray data.
\end{abstract}

\maketitle

Despite the fantastic success of the standard model of particle
physics (SM) and the standard model of cosmology, various fundamental
problems have accumulated that need attention. Many of these
problems stem from the ultraviolet (short distance) and infrared
(large distance) divergencies. There is a general consensus that
we understand our Universe (with some exceptions) on scales
approximately between $10^{-18}~{\rm m}$ and $10^{24}~{\rm m}$. The first scale
corresponds to the energy scale of TeV$^{-1},$ which is going
to be probed by the Large Hadron Collider (LHC); the second
scale corresponds to the isotropy scale of the Universe, {\it i.e.\/}, the distance at which cosmology truly takes over from astrophysics.

It is becoming increasingly clear that straightforward extensions of
existing theories either do not cure everything or bring in more problems. Some radically new ideas are needed to explain the Universe beyond these scales. In this Letter, we propose a novel approach introducing
the concept of evolving dimensionality: the effective dimensionality of
space depends on the energy scale we are probing and as the length
scale increases new dimensions open up. At short scales the space is
lower dimensional, at intermediate scales the space is
3D, and at large scales the space is effectively higher
dimensional. This set-up allows for some fundamental
problems in particle physics, gravity and cosmology to be addressed
from a new perspective. We will also show that this model can have striking
signals at the LHC.

This approach contrasts with attempts to solve the hierarchy problem by introducing extra spatial dimensions of a finite size, which open up at short
distances~\cite{ArkaniHamed:1998rs}. In our approach, at short distances spatial dimensions collapse and shut off one-by-one, resulting in a simple $(1+1)$ space-time at very short distances characteristic of the very early Universe right after the Big Bang, which has interesting implications for cosmology.  The approach is closer in spirit to DGP models and their cascading cosmology extensions~\cite{Dvali:2000hr}.

The approach proposed in this Letter is not a concrete model, but rather a conceptual new paradigm, which allows one to address known deficiencies of the standard models of both particle physics and cosmology in a radically new way. More concrete models within the proposed paradigm are required to resolve them explicitly.

One of the most acute problems connected with ultraviolet divergences
concerns radiative corrections to the mass appearing in
the Higgs potential, $V= -\mu^2 \Phi \Phi^\dagger + \lambda (\Phi^\dagger \Phi)^2$. The one-loop corrections to the Higgs mass from the top, $W,$ and Higgs self-coupling diagrams
are given by:
\begin{eqnarray} \label{terms}
i \frac{Y_t^2}{2}\int^\Lambda \frac{d^4k}{(2\pi)^4} \,
 {\rm tr} \left(\frac{i}{\not \! k -m_t}
\frac{i}{\not \! k+\not{\!p} -m_t} \right)
 &
\sim & \!\!\!
-\Lambda^2  \frac{Y_t^2}{8\pi^2}  \,, \nonumber \\
i \frac{g^2}{4} \int^\Lambda \frac{d^4k}{(2\pi)^4} \frac{1}{k^2-m_W^2} \sim  \Lambda^2 \, \frac{g^2}{64\pi^2}  \,, & ~&~~~ \\
i 6\, \lambda \int^\Lambda \frac{d^4k}{(2\pi)^4} \frac{1}{k^2-m_H^2}  \sim  \Lambda^2 \, \frac{3\lambda}{8\pi^2} \,, & ~& ~~~\nonumber
\end{eqnarray}
and grow quadratically. (The contribution from the $Z$ boson is
obtained with the substitutions $m_W \to m_Z$ and $g^2 \to g^2+{g'}^2.$) Here $Y_t$ is the top Yukawa coupling, $m^2_W = \frac{1}{4} g^2 v^2$,
$v = 246$~GeV, $m^2_Z = \frac{1}{4} (g^2 + g^{\prime 2}) v^2,$ $m_t^2
= \frac{1}{2} Y_t^2 v^2$, $m_H^2 = 2 \lambda v^2,$ $g$ and $g'$ are
the $SU(2)_L \times U(1)_Y$ gauge couplings, $\lambda$ is the quartic
Higgs coupling,
$\not{\!\!p}$ is the reduced four-momentum of the Higgs, and $\Lambda$ is an ultraviolet cutoff of the model, {\it i.e.\/}, the scale at which unknown physics beyond the SM takes over.

After summing over the color and polarization degrees of freedom one obtains~\cite{Veltman:1980mj}:
\begin{equation}
\Delta \mu^2 = \frac{3}{16 \pi^2 v^2} \, (2 m_W^2 + m_Z^2 + m_H^2 - 4 m_t^2)\, \Lambda^2 \, .
\end{equation}
Unless the Higgs mass is fine-tuned to an accuracy ${\cal
  O}(10^{-32})$, upon minimization of the potential, these corrections
result in a dangerous contribution to the Higgs vacuum
expectation value, which destabilizes the electroweak symmetry breaking scale. The SM
works amazingly well by fixing $\Lambda$ at the electroweak scale.  It is generally assumed
that this indicates the existence of new particles and laws of nature
at energies above $\Lambda$.  The alternative approach we exercise here is to keep  the structure of the SM and change the dimensionality of the background on which the SM
lives. A straightforward calculation shows that in 2D
all of the terms in (\ref{terms}) are only linearly divergent, {\it e.g.\/}, for the Higgs term:
\begin{equation}
i 6 \lambda \int^\Lambda \frac{d^3k}{(2\pi)^3} \frac{1}{k^2-m_H^2}  \sim   \Lambda \, \frac{3\lambda}{\pi^2},
\end{equation}
while in 1D all of these terms are only logarithmically divergent, {\it e.g.\/},
\begin{equation}
i 6 \lambda \int^\Lambda \frac{d^2k}{(2\pi)^2} \frac{1}{k^2-m_H^2}  \sim  \log(\Lambda/m_H) \, \frac{3\lambda}{\pi},
\end{equation}
thus alleviating the fine tuning problem. Indeed, from the dimensional regularization technique we know that ultraviolet divergences in field theory are associated with poles in the dimension plane. Therefore, lowering the dimensionality of space-time universally cures ultraviolet divergences in practically all of the field theories.

What about gravity? The most elusive concept in modern physics ---
the consistent quantization of  gravity --- is much more within the reach in lower
dimensions. Gravity in $(3+1)$ space-time is complicated, nonlinear, and perturbatively non-renormalizable theory.  However, if the fundamental short-scale physics is lower dimensional, there is no need to quantize $(3+1)$D gravity at short distances and we should quantize $(2+1)$ and $(1+1)$D gravity instead. In any space-time the curvature tensor $R_{\mu \nu \rho
  \sigma }$ may be decomposed into a Ricci scalar $R$, Ricci tensor
$R_{\mu \nu },$ and a conformally invariant Weyl tensor. In 2D the Weyl tensor vanishes and so
\begin{equation}
 R_{\mu \nu \rho \sigma } = \epsilon_{\mu \nu \alpha}
\epsilon_{\rho \sigma \beta} (R^{\alpha \beta} + \frac{1}{2} g^{\alpha
  \beta} R) \, .
\end{equation}
 This implies that any solution of the vacuum Einstein's equations, $R_{\mu \nu} =0$,
necessarily has a vanishing curvature, and consequently can be
constructed by ``gluing together'' flat pieces of Minkowski
spacetime~\cite{Staruszkiewicz:1963zz}.  This is most easily seen
after highlighting the precise equivalence between (2+1) vacuum
Einstein gravity and gauge theory~\cite{Achucarro:1987vz}. For a
spacetime manifold ${\cal M}$ of dimension three, the Einstein-Hilbert
action can be written in terms of the dreibein ($e^a = e_\mu^a
dx^\mu$) and the spin connection ($\omega^a = \frac{1}{2}
\epsilon^{abc} \omega_{\mu b c} dx^\mu$) as a Chern-Simons action
\begin{equation}
S = \frac{1}{8\pi G} \int_{\cal M} e^a \wedge \left( d\omega_a + \tfrac{1}{2} \epsilon_{abc} \omega^b \wedge \omega^c \right) \, .
\label{chern-simons}
\end{equation}
The equations of motion following from (\ref{chern-simons}),
\begin{equation}
T^a[e,\omega] = de^a + \epsilon^{abc} \omega_b \wedge e_c =0
\end{equation}
and
\begin{equation}
R^a[\omega] = d\omega^a + \tfrac{1}{2} \epsilon^{abc} \omega_b \wedge \omega_c =0,
\end{equation}
demand the spin connection to be flat and torsion free, and hence the curvature of the metric vanishes, $g_{\mu \nu} = e_\mu^a e_\nu^b \eta_{ab}$.  All in all, Einstein (2+1) dimensional vacuum gravity has no local degrees of freedom, {\it viz.}, there are no gravitational waves in the classical theory and no gravitons in the quantum theory. Therefore, if the high energy limit of spacetime is  described by a lower dimensional Einstein-Hilbert action, an striking prediction emerges: {\em the stochastic background of gravitational waves would have a horizon at the frequency scale where space transitions from 3D~$\rightleftharpoons$~2D.}  Interestingly, for a 3D~$\rightleftharpoons$~2D transition scale near 1~TeV, the horizon wavelength is within the reach of the future Laser Interferometer Space Antenna (LISA)~\cite{Mureika:2011bv}.

$(1+1)$D gravity is even more simple --- the gravitational coupling is dimensionless and the action is a topological invariant that gives no dynamics to the 2D metric. Interestingly, there is an asymptotically safe theory of pure gravity in ($2+\epsilon$) space-time dimensions ($\epsilon \ll1$); asymptotic safety can also be preserved in the presence of matter fields~\cite{Weinberg}. Dimensional continuation from $(2+\epsilon)$ to 4D is driven by the truncated exact RG equation~\cite{Reuter:1996cp}.  The RG flow predicts an effective dimensionality which is scale dependent: it equals 4 at macroscopic distances, but gets dynamically reduced at short distances and space-time becomes a 2D fractal~\cite{Lauscher:2001ya}.  Along the lines, the superposition of all possible Lorentzian spacetime shapes (a.k.a. causal dynamical triangulations) yields the same dynamical fractal structure~\cite{Ambjorn:2000dv}. However, it is noteworthy that the layered space-time structure envisaged in this Letter is not fractal, but has the properties of dimensions on the lattice.

\begin{figure}[htb]
\vspace*{-0.1in}
\scalebox{0.5}{\includegraphics{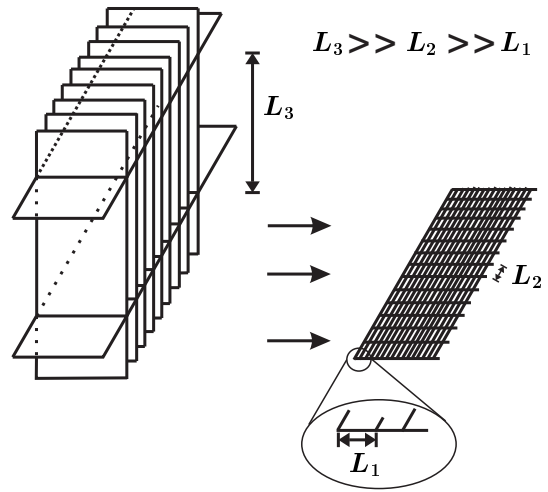}}
\vspace*{-0.1in}
\caption{Ordered lattice. The  fundamental quantization scale of space  is indicated by $L_1$. Space structure is 1D on scales  much shorter than $L_2 \simeq 4 \Lambda_2^{-1}$, while it appears effectively 2D on scales much larger than $L_2$ but  much shorter than $L_3 \simeq 4 \Lambda_3^{-1} $. At scales much larger than $L_3$, the structure appears effectively 3D. Following this hierarchy, at even larger scales, say $L_4$, yet another dimension opens up and the structure appears 4D (not shown in the picture).}
\label{lattice}
\end{figure}

Let us assume space-time has an ordered lattice structure, see Fig.~\ref{lattice}, which becomes anisotropic at very small distances. The proposed set up is analogous to that of dimensional crossover in layered strongly correlated metals~\cite{Valla}. These materials have an insulating character in the direction perpendicular to the layers at high temperatures but become metal-like at low temperatures, whereas transport within the layers remains metallic over the whole temperature range.

\begin{figure}[tpb]
\postscript{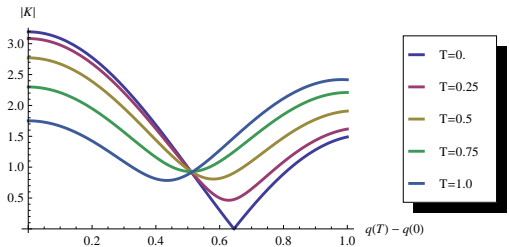}{0.8}
\caption{Probability distribution at different times, summing amplitudes
          from $n=1$ to $n=20000$. We have taken $T$ and $q(T)-q(0)$
          in units $L$, and $|K|$ in units of $A/L$.}
\label{probability}
\end{figure}

One can draw an instructive analogy between the standard path integral formalism and propagation of a particle on a lattice. In the path integral picture, a particle propagating from one to another point (separated by a macroscopic distance) does not always follow a straight path. Instead, due to quantum fluctuations, it follows an irregular jagged path. In fact, there are infinitely many possible paths that a particle can take, yet the straight classical trajectory and paths nearest to it give the highest contribution to the path integral. In the language of quantum mechanics, quantum interference of many possible paths gives a straight propagation on average. In the case of a particle propagating on the lattice, the geometry of the lattice, instead of quantum fluctuations, dictates the jaggedness of the path. The macroscopic straightness of the particle's path is maintained due to the initial momentum conservation. This path integral analogy can also help us visualize the dimensional reduction. Consider for example the phase-space path integral of a massless particle propagating over a 1D regular lattice of size $L$:
\begin{equation} \label{path} K\equiv \int {\cal D}p (t){\cal D} q (t)\exp i\int^T_0 dt(p\dot{q} -H(p,q)) \,, \end{equation} where $p(t)$ is the particle's momentum and $q(t)$ is the particle's location. In this 1D case there is no longitudinal space, the whole lattice is just a set of points separated by the distance L. The particle's allowed locations are points of the lattice, and the corresponding particle's wavelengths and momenta are $\lambda = 2nL$ and $p=2\pi/\lambda=\pi/(nL) ,$ where $n$ is a non-zero integer. Now take $\delta p = p(n)-p(n+1)= \pi/nL - \pi/(n+1)L \sim \pi/n^2L$
 so that $\int {\cal D} p$ can be replaced by $\displaystyle\sum_{n\in Z -\{0\}} \frac{A}{n^2 L}$, where $A$ is a normalization constant. If $T$ is very small, $\dot{q}$ can be treated as a constant and $q(T)=q(0)+\dot{q}T$;  Eq.~(\ref{path}) can be rewritten as
\begin{equation}
K=\sum_{n\in Z -\{0\}} \left[Ae^{ iT(H-p\dot{q})}\right]/\left[n^2L\right] \, .
\end{equation}
For a particle propagating from $(0,q(0))$ to $(T,q(T))$, the probability distribution is given by
\begin{equation}
|K|= \left|\sum_{n\in Z -\{0\}} \left[ Ae^{ i\pi\left(\frac{T}{|n|L}-\frac{q(T)-q(0)}{n L}\right)}\right] /\left[n^2L \right] \right| \,,
\end{equation}
where  $H=|p|=\frac{\pi}{|n|L}$. As can be seen in Fig.~\ref{probability}, the $|K|$ distribution has a peak near $(0, q(0))$ and a peak at $(L,q(0)+L)$. This implies that the particle most likely stays at $q(0)$ at small $T$ and jumps to $q(0) +L$ for $T \sim L$. For time scales which are a quarter of the lattice period  the peak of probability is at $q(0)$. This implies that  for energy scales $\Lambda \agt 4/L$ particles prefer to stay at the point (i.e. longitudinal space),  while low energy particles can propagate in the transverse space.

For $\Lambda_3 < \sqrt{s} < \Lambda_2$ the Universe is 2D and gravity, like any other force, is bound to 2D. (Note that this is not the case of bulk+branes, as here there is no bulk.) The world is truly 2D in a sense that the only third dimension is the thickness of the brane $\sim \Lambda_3^{-1}$ (i.e., the thickness of the spacing between lattice sheets).The Minkowski space-time metric shrinks to $(1,-1,-1).$ From Gauss's law, the gravitational potential becomes $\phi(r) = 2G_2 M \ln r$, where $G_2$ is the two dimensional gravitational constant.
The 3D~$\rightleftharpoons$~2D
 space transition takes place at a temperature $T_{_{3 {\rm D} \rightleftharpoons 2 {\rm D}}} \sim \Lambda_3 \agt 1~{\rm TeV}$, and so the earliest observationally verified landmarks ---~big bang nucleosynthesis (BBN) and cosmic microwave background (CMB) ---  stay unaffected as
\begin{equation}
T_{_{3 {\rm D} \rightleftharpoons 2 {\rm D} }} \gg   T_{_{\rm BBN}} \sim {\rm MeV} \gg T_{_{\rm CMB}} \sim {\rm eV}  \, .
\end{equation}

The transformation group acting on the lattice sheets, $SO(2,1)$, is a subgroup of the proper, orthochronous Lorentz group, $SO(3,1)$.  Generally one expects the lattice to fold randomly thus avoiding a creation of a preferred direction in space. Yet, there is a preferred ``cosmic'' reference frame in which the lattice is at rest. The local random orientation of the 2D substructure activates non-systematic violations of Lorentz symmetry in the low energy effective theory, i.e., for de Broglie wavelengths $\lambda > L_3$ (in the preferred frame of lattice coordinates).  The non-systematic effects induced a  local deformation of the dispersion relation, but the classical relation $E^2 = p^2 + m^2$ still holds on average~\cite{AmelinoCamelia:2002ws}.

The observed light from distant sources is continually subject to stochastic fluctuations of $L_3$, which introduce uncertainties in the determination of the photon wavelength $\delta \lambda \sim L_3 (\lambda/L_3)^ {1-\alpha}$, where $\alpha$ is a model dependent parameter that controls the cumulative effect of individual fluctuations. Since the fluctuations are uncorrelated, an initially in phase collection of photons will lose phase coherence as they propagate. For a propagation distance $L$, the cumulative statistical phase dispersion is $\Delta \phi \sim 2 \pi a L_3^\alpha L^{1-\alpha}/\lambda,$ where $a \sim 1$ is chosen from some model dependent probability distribution that reflects the underlying stochasticity~\cite{Ng:2003ag}. To constrain the parameter space, the strategy is to look for interference fringes for which the phase coherence of light from distance sources should have been lost, i.e., $\Delta \phi \agt 2 \pi$.  PKS1413+135, a galaxy at a distance of 1.2~Gpc that shows Airy rings at a wavelength of 1.6~$\mu$m~\cite{Perlman:2002fi}, is a typical probe. For $\Lambda_3 \sim 1~{\rm TeV}$, the allowed region of the parameter space, $\alpha \agt 0.8,$ encompasses the particularly interesting model of $\delta \lambda \sim L_3.$ This originates in non-renormalizable interactions suppressed by the mass scale characteristic of the lattice~\cite{Anchordoqui:2005gj}.

Fluctuations in the lattice structure would also spread a delta function type pulse of radiation in time. The overall time dispersion in the pulse can be related to the cumulative phase dispersion yielding $\Delta t \sim L_3^\alpha L^{1-\alpha} /v_\phi$, where $v_\phi$ is the phase velocity of the light wave~\cite{Ng:2003ag}.  The width of the pulse is not dispersive in frequency space. The expected time dispersion of a pulse originating at $3.6$~kpc is $\Delta t \sim 10^{12} \times (8 \times 10^{-39})^\alpha~{\rm s}$. High-precision timing observations of the pulsar B1937+21 (with  accuracies  $\sim 0.2\mu$s)~\cite{Kaspi:1994hp}, provide a somewhat weaker but compatible bound, $\alpha \agt 0.5$.

If in the preferred frame a de Broglie wavelength of a particle becomes significantly shorter than $L_3$, such a particle will propagate locally in 2D, rather than 3D.  Note that this does not affect the straightness of propagation of high-energy gamma rays from the source to the observer, as the overall momentum of the particle is preserved as it propagates through the spatial lattice. If the lattice is rigid enough, {\it i.e.\/}, the tension of the branes that form it is significantly higher than particle energy, the particle will scatter coherently at brane junctions and move along a jagged line preserving its original direction. This is similar to a photon propagating straight through a crystal lattice despite being scattered elastically off the individual atoms via phonon exchange. As long as the energy of the photon is small enough so that the scattering is elastic, the propagation of the electromagnetic wave through the crystal preserves the group velocity of the photon and its direction on the scales significantly larger than the lattice spacing. We can now apply this analogy to the scattering of a high-energy particle on the brane junctions in the lattice. The tension of the brane lattice and the relative sizes of the sides of its primary cell determine the refraction index. For $\lambda < L_3$, the particle propagates via a jagged trajectory with the degree of jaggedness given by the $L_2$ to $L_3$ ratio, which is the effective increase in the path length. For elastic interactions with the brane lattice, the refraction index for a high-energy particle becomes $1 + \Delta n$, where $\Delta n \sim L_2/L_3 \ll 1$. This implies that the dispersion relationship in the brane lattice is very non-linear and is characterized by a Fermi function with the threshold $\sim \Lambda_3 \sim 1$~TeV. The combination of this threshold-like behavior and the smallness of $\Delta t$ in the effective theory at long distances allows us to elude all the dispersion-like astrophysical constraints from TeV gamma rays~\cite{Aharonian:2008kz}.

For $\sqrt{s} > \Lambda_2 $~\cite{comment2}, when space becomes 1D, Minkowski metric is simply $(1,-1)$ and space and time in a sense become equivalent to each other.  Interestingly, if $CP$ was violated maximally in the Big Bang, then $T$ is also violated maximally (assuming the $CPT$ theorem still applies) and the fact that the time has a defined direction, while space does not, may simply come from that maximum violation (just as neutrinos are always left-handed due to the maximum violation of parity in weak interactions). Note that $CP$ maximally violated in the Big Bang, may also help to explain the observed extreme matter-antimatter asymmetry in the Universe.  (It is why the Big Bang would be completely $CP$-violating, of course, which remains to be explained in this picture.)

For distances $> L_4$, space becomes 4D, which would result in certain consequences for cosmology.  For example, if $L_4$ is of the order of the present cosmological horizon, a very small but finite positive cosmological constant can be attributed to the characteristic size of our 3D-lattice. Namely, the $(4+1)$-dimensional Einstein's equations admit the following metric as a vacuum solution
\begin{equation}
ds^2 = dt^2 - e^{2\sqrt{\Lambda/3}\, t} \left(dr^2+r^2d\Omega^2 \right) -
d\psi^2,
\end{equation}
where $\Lambda = 3/\psi^2$~\cite{PonceDeLeon:1988rg}. This metric
reduces on $\psi = {\rm const}$ hypersurfaces to a 3D de Sitter metric
with $\Lambda = {\rm const}$. An observer living on the 3D lattice (i.e., $\psi = {\rm const}$) will measure an effective stress energy tensor with the
equation of state $p=-\rho$, where $\rho = \Lambda \overline M_{\rm
  Pl}^2$ and $\overline M_{\rm Pl}$ is the reduced Planck mass.  Indeed, the observed vacuum energy density, $\rho \approx
(2.4 \times 10^{-3}~{\rm eV})^4$, corresponds to $\psi \approx 10^{61} M_{\rm
  Pl}^{-1} \approx 10^{26}~{\rm m}$. This is comparable to the current horizon size, which within the proposed scenario is comparable to the characteristic distance between 3D sheets comprising the 4D lattice structure.  If $L_4$ is of this scale, it represents the minimum value of $\psi$ which in turn represents the maximum (but not the minimum) value of the effective $\Lambda$ experienced by an observer on a 3D sheet.  It therefore sets the cosmological constant problem in a new context: as a Casimir energy due to the presence of another distant fold of the lattice.

Let us now focus attention on the lower dimensional crossover and its possible implications for LHC physics.  We discuss here three immediate and spectacular consequences of this model at the LHC, which should be observable if $\Lambda_3 \sim 1$~TeV, {\it i.e.\/}, within the reach of the machine: {\it (i)\/} cross section of hard scattering processes changes compared to that in the SM as the $Q^2$ becomes comparable with $\Lambda_3^2$; {\it (ii)\/} $2 \to 4$ and higher order scattering processes at high energies become planar, resulting, {\it e.g.\/}, in four-jet events, where all jets are produced in one plane in their center-of-mass frame, thus strikingly different from standard QCD multijet events; {\it (iii)\/} under certain conditions, jets of sufficiently high energy may become elliptic in shape.

Consider a $2 \to 2$ scattering in our brane-lattice model. If $Q^2$ of the scattering, {\it i.e.\/}, the degree of virtuality of the mediator (propagator) in the corresponding Feynman diagram becomes comparable to $\Lambda_3^2$, the mediating particle moves in 2D. The effective impact parameter of the interaction is now the impact parameter in the 2D plane defined by the local lattice geometry, hence the effective suppression of the cross section by the reduced phase space (and also modified matrix element of a 2D interaction). Since the spacing of the lattice $L_2$ is microscopic, different collisions will happen very large distance apart in the lattice space, and if the lattice is oriented randomly, the cross section will be suppressed  compared to the 3D case; with a more regular lattice orientation, the suppression factor could be different and even depend whether the collision took place in ATLAS or CMS or where in the lattice frame the Earth has been located at the time of the collision. (Such a time-dependent analysis is in principle possible, as the collision time is known to high precision.) In addition to this purely geometrical factor it is easy to see from dimensional arguments that the energy behavior of parton-level cross sections for 2D-scattering changes compared to that in 3D. For instance, the Drell-Yan cross section will drop not as $1/E^2$, but as $1/E^3$ once the 3D $\to$ 2D crossover energy is surpassed. The fact that this phenomenon has not been observed in the previous low-energy measurements, {\it e.g.,\/} at LEP and the Tevatron, can be interpreted as the bounds on the sharpness of the 3D $\to$ 2D crossover. Note that the fact that the propagator is bound to 2D, while the incoming particles move in 3D on the distances much greater than $L_2$ does not result in $T$ and $CPT$-violation, as the outgoing particles also propagate in 3D over long distances, due to the lattice back-reaction, which absorbs the momentum of the incoming particles in the direction perpendicular to the local 2D fold and then reemit it by giving the outgoing particles equal boost in the same direction.

Let's move on to the $2 \to 4$ scattering, which involves several virtual particles. If $Q^2$ in each of the propagators is comparable with $\Lambda_3^2$, the spatial separation between the incoming and outgoing particles at the time of the interaction is comparable to the size of the lattice. Thus, all the virtual particles (propagators) must move in the same 2D space transverse to the third dimension of the lattice, $L_3$. This results in the outgoing four partons to be in the same plane in the c.o.m. frame of the collision, thus drastically different from the 3D scattering, where four outgoing partons are in general acoplanar. As discussed above, the entire c.o.m. frame is boosted to conserve the longitudinal momentum of the incoming partons in the direction of the beam, but that does not affect the initial planar configuration, per the argument of photon propagation through the lattice. Thus, we expect, {\it e.g.\/}, multijet events with four or more jets at very high transverse momentum to become more and more planar as the characteristic $Q^2$ approaches $\Lambda_3^2 \sim 1$~TeV$^{2}$. The LHC sensitivity for identifying four jets coplanarities has been studied elsewhere~\cite{Anchordoqui:2010hi}.

Finally, if the lattice structure is similar over large distances (which is is generally not the case, as the lattice surfaces forming it may be folded and twisted in a non-trivial way), {\it i.e.}, over the distances comparable to $1/\Lambda_{\rm QCD}$, individual jets at very high energy may become elliptic in shape. This is due to the nature of the parton shower, which is generally ordered in $Q^2$; thus one expects the largest $Q^2$'s to happen at the beginning of the shower evolution. If several successive shower splittings have $Q^2 \sim \Lambda_3^2$ and the lattice orientation is preserved over the distance scale of the shower development, just like the multijet events become planar, the core of the jet will become planar as well. After the soft part of the parton shower is finished, the resulting jets will be elliptic rather than round in shape. It is not clear if this ellipticity will be large enough to be observable at the LHC, particularly given the fluctuations of parton shower within individual jets; nevertheless we believe that looking at the individual jet ellipticity as a function of jet energy may become an interesting experimental probe of models with vanishing dimensions. In fact these jets may have been already observed by the Pamir Collaboration in showers induced by high-energy cosmic rays (the effect know as ``alignment'')~\cite{Pamir}, which can not be explained by conventional physics.

With the LHC just having achieved high-energy collisions at the half of the design energy and preparing for a decade of exciting explorations at the Terascale, it's very timely to bring the paradigm of vanishing dimensions to the attention of the experimental and theoretical communities, which is the main goal of this Letter

\section*{Acknowledgements}

This work is partially supported by the US National Science Foundation, under Grants No. PHY-0757598, PHY-0914893, and CAREER PHY-1053663, US Department of Energy, under Grant No. DE-FG02-91ER40688, and EU Marie Curie Network UniverseNet (HPRN-CT-2006-035863). The authors would like to thank the organizers of the ``Extra Dimensions and Mini Black Holes'' Workshop in Heidelberg, July 24--25, 2009, where the original idea of this article was conceived.


\begin{thebibliography}{9}
\bibitem{comment2} For $10~{\rm TeV} < \Lambda_2 < 1~{\rm PeV}$, the fine-tuning of the Higgs mass is between 10\% and 0.1\%.
\end{thebibliography}
\end{document}